\documentclass[10pt]{article}
\usepackage{fleqn,multicol,graphicx}
\begin{document}
\begin{titlepage}
\begin{center}
{\Large\sffamily\bfseries 
Expected neutrino signal from supernova remnant \\[.5ex]
RX J1713.7-3946 and flavor oscillations 
}
\end{center}

\vspace{4mm}
\centerline{\large Maria Laura Costantini$^{a}$ and Francesco Vissani$^b$}

\centerline{\em $^a$ Universit\'a dell'Aquila, L'Aquila, Italia and INFN}
\centerline{\em $^b$ INFN, Laboratori Nazionali del Gran Sasso,
Assergi (AQ), Italia}
\vspace{6mm}

\centerline{\large  Abstract}
\begin{quote}
\small 
We consider the impact of oscillations on 1-200~TeV 
neutrinos expected from RX~J1713.7-3946.
After a description of the nature of the source,
we obtain a prediction for the neutrino fluxes, 
based on the intense gamma ray flux first seen by CANGAROO and 
recently measured by H.E.S.S.\ experiment.
We study the effect of 3 flavor oscillations in detail and 
consider the impact on the muon flux induced by these high energy neutrinos,
potentially observable by a neutrino telescope located in the 
Northern hemisphere. A detector in the Mediterranean 
with an effective area of 1 km$^2$ and unit detection efficiency
should be able to see a signal of about 10
muons per year.

\end{quote}
\rm 
\end{titlepage}
\begin{twocolumn}


\section*{Introduction\label{cbd}}
The search for high energy cosmic neutrinos has a long history, 
tightly connected with the history of cosmic rays.
The hope of a happy end is linked to neutrino 
telescopes presently in operation, in construction or in project.
In the present work, we focus on a promising potential galactic source,
the young supernova remnant RX J1713.7-3946, 
and calculate the expected muon signal
using standard techniques \cite{b,gai}.

We begin by a description of this object (sect.~\ref{obj}), and 
state the predicted neutrino flux in sect.~\ref{nf}.
In sect.~\ref{osc}, we evaluate precisely 
the impact of three flavor oscillations on the flux.
The effect of the 
absorption of the high energy part of the flux in the Earth 
and of the live-time of observation 
is considered in sect.~\ref{abs}. The muon signal in
neutrino telescopes is studied in 
sect.~\ref{snt}. 
A summary of the result is 
offered in sect.~\ref{dis}.

\section{A potential neutrino source\label{obj}}
In this section, we recall the motivations 
of interest in supernova
remnants, and a number of recent facts that 
suggest that a specific SNR, RX J1713.7-3946 (G347.3-0.5),
is a promising source of high energy neutrinos.

SNR are likely to act as accelerators
of cosmic rays (CR). This suspicion was raised in 1934 \cite{baa}
and convincingly supported thirty year later \cite{gs}
on the basis of energetics:
if several percents of the energy injected by one supernova
(${\cal E}\sim 10^{51}$ erg) go in CR acceleration, 
the losses from the Galaxy can be compensated.
Two monographs, appeared in 1984 and 1990,
offer still very actual summaries of 
the astrophysics of cosmic rays  \cite{b} and the 
connections with particle physics \cite{gai}.
To a certain extent, the theory of acceleration of CR in SNR 
is still in evolution, but the generic expectations are stable:
the cosmic ray flux at the SNR is 
expected to be a power law spectrum:
\begin{equation}
F_p= K\cdot E^{-\Gamma}
\label{eq1}
\end{equation}
with index $ \Gamma=2.0-2.4 $ and maximal energy 
$E_{p,max}$ possibly as large as several PeV,
as suggested by `knee' of the CR spectrum
seen with extensive air showers arrays.
Due to galactic magnetic fields,
we cannot trace back CR to their source directly, 
but we can reveal sources 
if CR interact with some dense target
(cosmic beam-dumps).
The CR  would partially fill the dense region, 
their interactions would produce mesons, which would eventually 
decay yielding observable gamma and neutrino radiation.
The best case is a molecular 
cloud near to a young SNR~\cite{aha1,aha2,naha}.

There are converging indications that this  
happens in one specific SNR visible 
in the Southern sky, RX J1713.7-3946.
Let us recall the main points:
$(i)$~A strong X ray source has been discovered 
there \cite{x}, that is compatible with 
a core collapse SN exploded in A.D.~393 at a 
distance of about 1 kpc~\cite{wang}.
$(ii)$~The SNR is probably interacting with a 
molecular cloud, at about the same distance  
\cite{nanten}. This was observed through the 
CO molecule; 21 cm hydrogen line observations corroborate
these indications \cite{koo}.
$(iii)$~A large portion  of the X-radiation comes 
from the same region where the cloud happens to be, 
the column density that produces X-ray absorption
is compatible with the observed molecular cloud, and interesting
details are continuing to emerge~\cite{x1,x2,x3,x4,x5,x6,vvv}. 
The interpretation of the X-rays as synchrotron radiation 
from 100 TeV electrons is compatible with ATCA radio observations.
$(iv)$~But most interestingly, the CANGAROO team did observe 
TeV gamma rays since several years \cite{CA1}. 
This suggested this source as a CR accelerator \cite{butt1}.
Later it was claimed \cite{CA2} that the only 
likely mechanism to produce the bulk of 
gammas  is the hadronic one (namely, $\pi^0$'s from proton interactions). 
Recently, the H.E.S.S.~experiment also reported on the observation of 
an intense source of gamma rays, with energy in 
the TeV-10 TeV energy range \cite{hess}. This adds support 
to the overall picture, and (already with the first data) 
offers a precise determination of the photon flux.

One could perhaps argue that none of the items
discussed above, taken alone, seems to be conclusive.
Also, the interpretation outlined here
has to face a number of controversial points:
the distance of the object \cite{slane}, 
the compatibility with EGRET bound \cite{pohl}
(but see \cite{tanimori})
and the uniqueness of the hadronic hypothesis \cite{butt}
have been all questioned. 
Furthermore, the spectra of CANGAROO and H.E.S.S.\ do not agree well 
(see \cite{hess}); this may be an indication that 
the systematic error for energy measurement of one or both experiments 
has an underestimated uncertainty 
(note however that CANGAROO 
has measured photons from  the N-W rim, while H.E.S.S.\ measures the spectrum
for the photons coming from a wider region).
All these objections have to be seriously considered. However,
the fact remains that  RX~J1713-3946 is a very promising 
case for a cosmic beam dump, where the available 
observations seem to meet theoretical expectations.
In this case, the observed gamma radiation must be 
accompanied by neutrinos. In view of the interest 
in neutrino telescopes located the Northern hemisphere,
this is a very important conclusion. 
Actually, we would dare to say that RX J1713-3946 is at present 
the most definite hope (although not necessarily `the best') 
of a successful observation of cosmic neutrinos. 
For other possible sources of TeV neutrinos 
in the Galaxy, see~\cite{teresa}.

As a matter of fact, 
there is already a specific calculation of the neutrino 
signal from this source \cite{amh}. 
We improve on this calculation in the following points: 
we consider deviations from the hypothesis $\Gamma=2$, 
we include the effect of live-time of measurement and of
neutrino absorption, we describe the interactions at 
next-to-leading order (NLO) in QCD,
and most importantly, we consider 
the occurrence of neutrino oscillations.

\section{Secondary gamma and neutrino radiation\label{nf}}

The connection between gamma and neutrino is described in 
\cite{b} (see in particular ref.[38] of Chapter VIII there,
or tab.1 of~\cite{altr}) and in~\cite{gai}.
Here, we will follow this last reference quite closely, and
describe the relation between secondary gamma and 
neutrino radiation using the formul\ae\ 
of cascade theory.\footnote{The situation 
is simpler than for the Earth
atmosphere, since the muons originating 
from the leptonic decays of 
charged pions or kaons do not interact. 
The same is true for the photons 
from neutral pion and eta decays, since the molecular 
cloud is very thin in comparison with the radiation length 
$X_0\sim 60\ \rm g/cm^2$ for hydrogen
(=the source is ``gamma-transparent'').
Indeed, a cloud of a few parsec and 
with at density of few times 100 particles/cm$^3$ has 
a column density around 0.005~g/cm$^2$ or smaller.} 
Assuming scaling, CR primaries and secondaries 
(photons, neutrinos and antineutrinos) have 
the same type of spectrum. 
So we take as injection proton spectrum  
a power law with spectral index $\Gamma$ in the range $2-2.4$
as in eq.~(\ref{eq1}), and similarly for neutrinos.
The photon spectrum
from the  cascade $p\to \pi^0\to \gamma$ is 
\begin{equation}
F_\gamma= \frac{\Delta X}{\lambda_p}\cdot 
\frac{Z_{p\pi^0}(\Gamma)}{\Gamma}\cdot F_p
\end{equation}
where $\Delta X$ is the column density traversed by the protons
and $\lambda_p$ is the interaction length of CR. 
The effects of the $\pi^0$ distribution (determined
by strong interactions) are lumped into the 
spectrum-weighted momenta, $Z_{p\pi^0}$ in this example.
Similar expressions hold for neutrinos, as a sum of several 
(slightly more complicated) terms that describe the 
possible branches of the $\pi$ and $K$ cascades.\footnote{
For instance, since a $\nu_e$ comes from  $p\to \pi^+\to \mu^+\to \nu_e$,
we have 
$F_{\nu_e}= {\Delta X}/{\lambda_p} \cdot 
Z_{p\pi^+}(\Gamma) \cdot f_{\pi^+\mu^+ \nu_e}(\Gamma) 
\cdot F_p + ...$, where $f_{\pi^+\mu^+ \nu_e}$ is a function
that can be found in Sect.~7.1 of the book \cite{gai}, 
and `...' stands for the additional $K^+$ contribution, weighted with the 
branching ratio 
into leptons, $BR=0.635$.
Values of the spectrum-weighted momenta are obtained 
from fig.~5.5 of the same book; in this way, we introduce an error  
at the few \% level.}
In summary, the flux of the 
neutrinos of any species is just proportional to the photon flux:
\begin{equation}
F_{\nu}=k \cdot F_\gamma, \ \ \ \ 
\nu=\nu_\mu,\bar{\nu}_\mu,\nu_e,\bar{\nu}_e
\end{equation}
through a proportionality coefficient $k$ 
that depends on the type of neutrino and on the 
spectral index. Numerical values are given in table~\ref{tab1}.
In this manner, the neutrino fluxes are {\em predicted} 
in terms of the measured photon flux.

H.E.S.S.~data \cite{hess} in the range $E=1-10$~TeV
are well described by a
power spectrum with $\Gamma=2.19\pm  0.09\pm 0.15$,
that is in good agreement with theoretical expectations.
For our purposes, and since these measurements are 
going to be improved soon in the future, 
we shall limit ourselves to set $\Gamma=2.2$, modeling 
the photon flux as follows:
\begin{equation}
\begin{array}{l}
F_\gamma=1.7\times 10^{-11}\ 
\left(\frac{E}{\rm TeV}\right)^{-2.2}
\frac{1}{\rm TeV cm^{2} s }\\[1ex]
\end{array}
\end{equation}
The corresponding neutrinos fluxes are:
\begin{equation}
\begin{array}{l}
F_{\nu_\mu}^0=7.3\times 10^{-12}\ 
\left(\frac{E}{\rm TeV}\right)^{-2.2}
\frac{1}{\rm TeV cm^{2} s }\\[1ex]
F_{\overline{\nu}_\mu}^0=7.4\times 10^{-12}\ 
\left(\frac{E}{\rm TeV}\right)^{-2.2}
\frac{1}{\rm TeV cm^{2} s }\\[1ex]
F_{\nu_e}^0=4.7\times 10^{-12}\ 
\left(\frac{E}{\rm TeV}\right)^{-2.2}
\frac{1}{\rm TeV cm^{2} s }\\[1ex]
F_{\overline{\nu}_e}^0=3.0\times 10^{-12}\ 
\left(\frac{E}{\rm TeV}\right)^{-2.2}
\frac{1}{\rm TeV cm^{2} s }
\label{flussi0}
\end{array}
\end{equation}
where the superscripts $^0$ remind us that oscillations are not
included. In this approximation, the
flux of tau (anti) neutrinos at the source
is expected to be negligible.

\begin{table}[t]
\begin{center}
\begin{tabular}{|c|cccc|}
\hline
spectr.\ index 
& $\nu_\mu/\gamma$ & $\overline{\nu}_\mu/\gamma$  & $\nu_e/\gamma$ & 
$\overline{\nu}_e/\gamma$ \\[.5ex]
\hline\hline
2.0& 0.50 & 0.50  & 0.30  & 0.22  \cr
2.1& 0.46 & 0.46  & 0.29  & 0.19  \cr
2.2& 0.43 & 0.43  & 0.28  & 0.18  \cr
2.3& 0.40 & 0.41  & 0.26  & 0.16  \cr
2.4& 0.37 & 0.38  & 0.25  & 0.15  \cr
\hline
\end{tabular}
\end{center}
\vskip-3mm
\caption{\em Fluxes of neutrinos and antineutrinos originating
from pion and kaon decay, relative to the photon flux.
The latter is assumed to be a power spectrum with value of index
indicated in the $1^{st}$ column.\label{tab1}}
\end{table}

An important question is which is the uncertainty on the
neutrino/photon ratio. A primary cause is
the uncertainty in the photon flux from $\pi^0$s. 
Beside experimental errors, it is possible that 
the gamma radiation has other, non-hadronic 
components; this will be better quantified with more data and 
when the morphology of the source will be understood in detail.
The uncertainties in column density $\Delta X$ 
disappear when we consider the ratio.
Other causes of uncertainty include the one on 
hadronic interactions (=the spectrum-weighted 
momenta) and the neglected decay channels, 
but again this should have a weaker impact on the ratio.
If we consider as an analogy 
the predictions of the atmospheric neutrino fluxes 
\cite{atma}, 
we are lead to believe that the neutrino fluxes we deduced should 
have an accuracy of 20\% or better,
at least in the energy region
of $1-10$~TeV. Another way to argue for such an
accuracy  is to compare the results
of our tab.\ref{tab1} with those in \cite{altr}.
Now, if one agrees that an accuracy of 20~\% is reached, 
the effects of oscillations {\em must be included},
since as we will see they are of the order of~50\%.

\section{Three flavor oscillations of SNR neutrinos\label{osc}}
Now, we pass to describe the effects of 
neutrino oscillations.
{}From the theoretical point of view,
the situation is particularly simple,
since the phases of oscillations 
are really very large: 
\begin{equation}
\varphi\sim 
3\cdot 10^8
\left(
\frac{\Delta m^2}{8\cdot 10^{-5}\ \rm{eV}^2}
\right)
\left(
\frac{D}{1\; \rm{kpc}}
\right)
\left(
\frac{10\; \rm{TeV}}{E_\nu}
\right)
\end{equation}
The conclusion is that we just need 
to consider averaged vacuum\footnote{The 
MSW \cite{msw} effect does not modify the conclusion 
for two different reasons: 
(1)~in the vicinity of the star,
the matter potential is negligible
in comparison to the vacuum term
because of the small density 
in the molecular cloud;
(2)~inside the Earth, the converse happens;
the matter potential is so large that 
any further oscillation is suppressed. 
See also \cite{alyo}.} 
oscillations~\cite{pont,crock1,crock2,ath,bbhpw}.
The expression of the probability of flavor transformation
is given in function of the mixing matrix $U_{\ell j}$: 
\begin{equation}
P_{\ell\ell'}=\sum_j |U_{\ell j}^2|\cdot |U_{\ell' j}^2|,
\end{equation}
with $\ell,\ell'=e,\mu,\tau$.
The probabilities are the same for neutrinos and antineutrinos.
After propagation, the neutrino fluxes become:
\begin{equation}
F_\ell=\sum_{\ell'=e,\mu,\tau} P_{\ell\ell'}\; F^0_{\ell'}
\end{equation}

Adopting the standard decomposition of  $U_{\ell j}$ \cite{pdg},
we can summarize the present experimental information as:
$\theta_{12}=32.5^\circ \pm 2^\circ$ (solar neutrinos and KamLAND),
$\theta_{23}=45^\circ \pm 10^\circ$ (atmospheric neutrinos and K2K),
$\theta_{13}=0^\circ \pm 10^\circ$ (CHOOZ), 
$\delta_{\textrm{\tiny{CP}}}=0^\circ-360^\circ$ (namely, we do not know the CP violating phase 
$\delta_{\textrm{\tiny{CP}}}$, but it appears always with $\theta_{13}$). 
In a reasonable approximation, 
the symmetric matrix $P$ (with elements $P_{\ell\ell'}$) 
is given by:
\begin{equation}
P\sim 
\left(
\begin{array}{ccc}
0.6 & 0.2 & 0.2\\
& 0.4 & 0.4 \\
& & 0.4 
\end{array}
\right)
\end{equation}

An interesting question is which 
deviations we can expect. Let us assume that there are not main 
causes of systematic errors. Since the formal 
errors in the angles are quite small, it is useful to expand in linear
approximation in $\theta_{12}$, $\cos2\theta_{23}$ and $\theta_{13}$
around the central point 
$\theta_{12}=32.5^\circ$, 
$\cos2\theta_{23}=0$ and
$\theta_{13}=0^\circ$, getting:
\begin{equation}
P\simeq
\left(
\begin{array}{ccc}
1-\frac{x}{2} & \frac{x}{4}+y & \frac{x}{4}-y \\
 & \frac{1}{2}-\frac{x}{8}-y & \frac{1}{2}-\frac{x}{8} \\
 & & \frac{1}{2}-\frac{x}{8}+y 
\end{array}
\right) 
\end{equation}
where we 
define $x=\sin^2 2\theta_{12}$,  $y=\epsilon_{23}+\epsilon_{13}$ 
and:
\begin{equation}
\left\{
\begin{array}{l}
\epsilon_{12}=2 \sqrt{x (1-x)}\cdot \delta \theta_{12}\\
\epsilon_{23}=x/4\cdot \cos2\theta_{23}\\
\epsilon_{13}=\sqrt{x (1-x)}/2\cdot \delta \theta_{13}\cdot \cos\delta_{\textrm{\tiny{CP}}}
\end{array}
\right.
\end{equation}
Thus, the three uncertainty in  $P$, respectively 
due to the angles $\theta_{12}$, $\theta_{23}$ 
and $\delta_{\textrm{\tiny{CP}}}$ (setting $\delta \theta_{13}=10^\circ$),  
are:
\begin{equation}
\begin{array}{l}
\delta P\simeq
\pm 2.7\% 
\left(
\begin{array}{rrr}
-1 & 1/2 & 1/2\\
& -1/4 & -1/4 \\
& & -1/4 
\end{array}
\right)
+\\
+ (\ \pm 3.6\%\ \pm 3.3\%\ )
\left(
\begin{array}{rrr}
0 & 1 & -1\\
& -1 & 0 \\
& & 1 
\end{array}
\right)
\end{array}
\end{equation}
{}From previous equation we see that the variations are rather small.
The main effect when we are interested to muon signal is due to
the latter two uncertainties. Combining them in quadrature we obtain
the numerical expression:
\begin{equation}
\delta P_{\mu\mu}=-\delta P_{e \mu}=\pm 5 \%
\end{equation}
which means that the errors introduced by the 
uncertainties in the parameters of oscillations 
are negligible. (It means also that there is little
hope to learn anything useful on 3 flavor oscillations). 

Thus we evaluate oscillations with mixing angles at central values.
Using the fluxes in eq.~(\ref{flussi0}), 
we arrive at the following expectation
for neutrino fluxes at Earth:
\begin{equation}
\begin{array}{l}
F_{\nu_\mu}=3.9\times 10^{-12}\ 
\left(\frac{E}{\rm TeV}\right)^{-2.2}
\frac{1}{\rm TeV cm^{2} s }\\[1ex]
F_{\overline{\nu}_\mu}=3.5\times 10^{-12}\ 
\left(\frac{E}{\rm TeV}\right)^{-2.2}
\frac{1}{\rm TeV cm^{2} s }\\[1ex]
F_{\nu_e}=4.3\times 10^{-12}\ 
\left(\frac{E}{\rm TeV}\right)^{-2.2}
\frac{1}{\rm TeV cm^{2} s }\\[1ex]
F_{\overline{\nu}_e}=3.3\times 10^{-12}\ 
\left(\frac{E}{\rm TeV}\right)^{-2.2}
\frac{1}{\rm TeV cm^{2} s }
\label{flussi}
\end{array}
\end{equation}
They are the same within 20\%.
The expected flux of tau (anti) neutrinos is the same 
as the flux of muon (anti) neutrinos, which could lead 
to interesting signals.
However, in the following we focus just 
on the {\em muon} neutrino and antineutrino fluxes. They give rise to 
muons, thus offering a simple way to emphasize an observable 
signal.

\section{Live-time and absorption in the Earth\label{abs}}
During the sidereal day 
which lasts $2\times \tau=23^h 56^m 4^s$, 
a neutrino telescope can observe a source 
only when the overwhelming background atmospheric muons 
is absent. In first approximation, this condition is met
when the source is below the horizon. 
Taking the Earth's axis of rotation
as $\hat{z}$ direction, and $\hat{x}$ axis in such a manner that the 
source is in the $xz$ plane, the direction of the source is 
$\hat{s}=(\cos\delta,0,\sin\delta)$ 
($\delta$ is the declination, $\delta=-39^\circ 46'$ in our case)
and the one of the telescope is 
$\hat{t}=(\cos\phi \cos(\pi t/\tau),
\cos\phi \cos(\pi t/\tau),
\sin\phi)$ 
($\phi$ is the latitude).
ANTARES has $\phi=42^\circ 50'$ (that we adopt 
for numerical example), 
NEMO or NESTOR are more South, about 
$\phi= 36^\circ 30'$ and $\phi= 37^\circ 33'$ respectively, 
whereas BAIKAL is more North $\phi =51^\circ 50'$.
The cosine of the zenith angle 
$\cos\theta_Z\equiv \hat{s}\cdot \hat{t}$ is thus:
\begin{equation}
\cos\theta_Z=\sin\delta\sin\phi+\cos\delta\cos\phi \cos(\pi t/\tau)
\label{ccc}
\end{equation}
The origin of the time $t=0$ is the point of highest 
altitude (the apex), and conversely, the lowest altitude is reached
when $t=\tau$. The source becomes observable 
after the time $\tau_0$ that satisfies
$\cos\theta_Z(\tau_0)=0$.
This condition can be satisfied 
if $90^\circ-|\delta|\ge \phi \ge -(90^\circ-|\delta|)$,
that happens to be true for all detectors except BAIKAL.
In other words, RX J1713-3946 is always observable for 
BAIKAL (it is always below the horizon, $\phi-\delta>90^\circ$),
whereas for the other detectors, it is observable 
for a fraction of time $f_{liv}=1-{\tau_0}/{\tau}$.
This can be written:
\begin{equation}
f_{liv}=1-\frac{\arccos(-\tan\delta\; \tan\phi)}{\pi}
\end{equation}
For ANTARES this is 78~\%, whereas
for NEMO and NESTOR this is a bit less, 71~\% and 72~\% respectively.
\begin{figure}[t]
\includegraphics[width=.35\textwidth,angle=270]{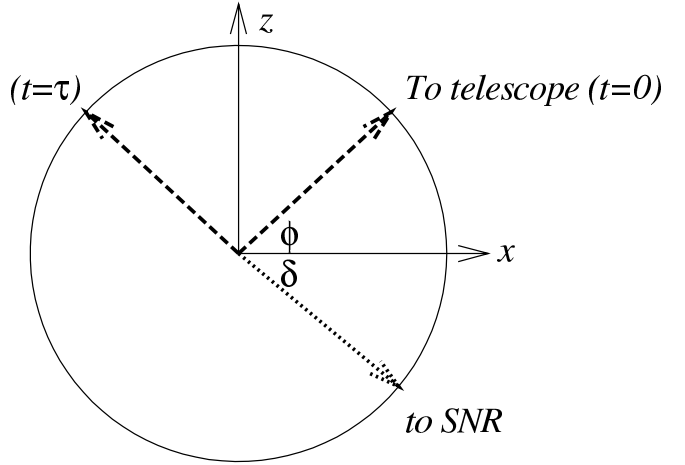}
\caption{\em Projection of the Earth in the $xz$-plane. The versor
of the SNR (with direction ``to SNR'')
is shown. Also shown the versor of the telescope 
at the times $t=0$ and $t=\tau$, 
when it lies in the $xz$-plane. \label{fig0}}
\end{figure}

There is another effect that 
diminishes the number of observable events:
High energy neutrinos are absorbed in the Earth 
before reaching the detector. 
This effect depends on
the column density $x$ seen by neutrinos.
When $\cos\theta_Z\le 0$, we have:
\begin{equation}
x=-2 R_\oplus \cos\theta_Z\cdot \overline{\rho}_\oplus(\cos\theta_Z)
\end{equation}
This varies with time according to eq.~(\ref{ccc}). Here, 
$R_\oplus=6.371\cdot 10^8$~cm is the radius of the Earth, 
and $\overline{\rho}_\oplus$ (in gr/cm$^3$) 
is the average Earth density along the 
line of sight, obtained using the PREM model \cite{prem}.
Now we can define the neutrino 
absorption coefficient $a_\nu$ and its 
time average $\overline{a}_\nu$
as:
\begin{equation}
\begin{array}{l}
a_\nu(t,E)=1-e^{-N_A x(t)\sigma(E) }\\
\overline{a}_\nu(E)=\frac{\int_{\tau_0}^\tau dt\ a_\nu(t,E) }{\tau-\tau_0}
\end{array}
\label{asso}
\end{equation} 
where $N_A$ is the Avogadro number and $\sigma$ the 
effective cross section of neutrino absorption.
The main part is due to CC interactions; 
NC interactions increase the absorption 
coefficient by a small amount.\footnote{The NC cross section is 
$\sim 1/3$ of the CC one for relevant energies. At first,
one could guess that $\sigma=\sigma_{cc}+\sigma_{nc}\equiv\sigma_{tot}$.
But NC interactions differ from CC interactions since they do not
absorb a neutrino; rather, they lower its energy (`neutrino regeneration').
We can estimate this effect by 
replacing $\sigma_{nc}$ with $\sigma_\Gamma\equiv\sigma_{nc}(1-Z_\Gamma)$, 
where setting $E'=E/(1-y)$ we define
$Z_\Gamma(E)\ \sigma_{nc}(E) \equiv
\int dy\; (1-y)^{(\Gamma-1)} \; d\sigma_{nc}(E',y)/dy$ 
\cite{ginz}. 
For $\Gamma\approx 2.2$, $\sigma_\Gamma$ is about 10~\% 
of $\sigma_{cc}$.
Using in the exponent of eq.~(\ref{asso}) 
$\sigma\equiv\sigma_{cc}+\sigma_\Gamma$ with $\Gamma=2.2$, we conclude that 
$\overline{a}_\nu(E)$ increases by about 5~\%.}
This conclusion is in agreement with what found in \cite{igt}.
For antineutrinos, the calculations are exactly the same. 

For the interactions (deep inelastic scattering)
we ad\-opt the recently calculated MRST2004 partons \cite{mrst}
and work at NLO in QCD \cite{mrstnlo}.
Let us note that the measurements of HERA at $s=4 E_p E_e$
$\equiv  2 M_p E_\nu$ with $E_\nu\sim 50$~TeV give us confidence 
that we have an accurate description of the interaction 
cross section in the relevant energy range.
Small modifications due to nuclear 
medium (average rock nuclei in this section,
water nuclei in the next one)
are described using the simple prescriptions of~\cite{smi}.

\section{Signal in neutrino telescopes\label{snt}}
Charged-current interactions of $\nu_\mu$ and $\overline{\nu}_\mu$
produce muons and antimuons that can be observed 
underground. This is the simplest observable for a cosmic source
of high-energy neutrinos, and it is known since long \cite{markov}. The reason
why we prefer to concentrate on this observable is that 
the underwater detectors can achieve a very good angular resolution,
perhaps better than one degree. This offers a very effective tool  
to reject the background of atmospheric neutrinos in neutrino telescopes.

The number of muons and antimuons reaching an area $A$ in 
a time of observation $T$ is:
\begin{equation}
\begin{array}{l}
N_{\mu+\overline{\mu}}=\displaystyle f_{liv} \cdot A\cdot T\cdot
\int_{E_{th}}^{\infty} dE_\nu\; 
F_{\nu_\mu}(E_\nu)\times  \\[2ex]
\ \  \ \times Y_{\mu}(E_\nu,E_{th})
(\; 1-\overline{a}_{\nu_\mu}(E_\nu)\; )+ 
({\nu}_\mu \to \overline{\nu}_\mu)
\end{array}
\end{equation}
where $E_\nu$ is the energy of the neutrino at the point of interaction
and $E_{th}$ is the minimal muon energy that 
can be detected, and
``${\nu}_\mu \to \overline{\nu}_\mu$'' stands for the contribution of the 
antineutrinos (same expression using antineutrino flux, cross section, 
and absorption coefficient).
Whenever needed, we take as reference values:
\begin{equation}
A=1\mbox{ km}^2,\ T=1\mbox{ solar y},\ E_{th}=50\mbox{ GeV}
\end{equation}
Neutrino fluxes $F$ and absorption coefficients $\overline{a}_\nu$
are defined in eqs.~(\ref{flussi0},\ref{flussi}) 
and (\ref{asso}). Finally, the 
probability 
to yield a muon $Y_{\mu}$ 
can be calculated by the 
interaction cross sections and the muon range in water
in the  following manner:
\begin{equation}
Y_{\mu}=N_A \int_{E_{th}}^{E_\nu} \! dE_\mu 
\frac{d\sigma_{cc}}{dE_\mu}(E_\nu,E_\mu)\ R(E_\mu,E_{th})
\end{equation}
where $N_A$ is the Avogadro number; similarly for antineutrinos.
The muon range $R(E_\mu,E_{th})$ can be obtained integrating 
the equation: 
\begin{equation}
\frac{dR}{dE_\mu}=-\frac{1}{\alpha+\beta E_\mu}
\end{equation}
where the dependence of $\alpha$ and $\beta$ on $E_\mu$ 
in water is taken from 
ref.~\cite{b}.\footnote{We recall that 
in the approximation of constant 
coefficients,   
$R(E_\mu,E_{th})=1/\beta \log[(1+E_\mu/\epsilon)/(1+E_{th}/\epsilon)]$
with $\epsilon=\alpha/\beta$. 
In the energy range of interest this agrees at 10~\% 
with the accurate result when $\alpha=2.4\cdot 10^{-3}$ GeV/cm and 
$\beta=2\times 10^{-6}$~cm$^{-1}$.} 
On passing, we
remark that occasionally the calculation of the 
the cross sections $\sigma_{cc}$ and of the yields $Y_\mu $ 
are done using the DIS formula at the leading order (LO), but using 
the partons calculated at NLO. This procedure is not consistent, 
and the cross sections and yields obtained in
this way are overestimated by 10\% at 20 TeV,
and by 25\% at 1 PeV.

\begin{figure}[t]
\includegraphics[width=.44\textwidth,angle=270]{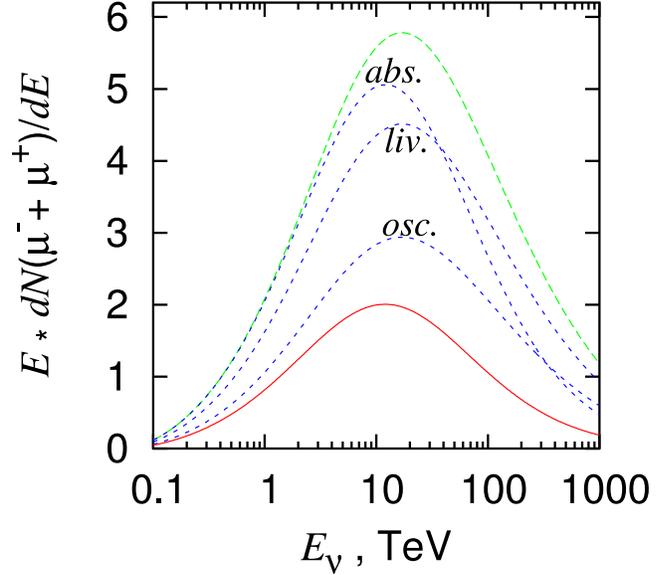}
\caption{\em Distribution of muons + antimuons 
above $E_{th}=50$~GeV
due the fluxes of eqs.~(\ref{flussi0}) and (\ref{flussi}), 
that pass through an area $A=1$~km$^2$ in 1~yr.
The higher curve is the `idealized' case.
The 3 middle ones show the 
impact of absorption (``abs.''), live-time (``liv.'')
and 3 flavor oscillations (``osc.''). 
The last effect is universal,
whereas the first 2 effects are estimated for $\phi=42^\circ 50'$.
The lower curve includes all three effects.\label{fig1}}
\end{figure}

The parent spectrum (distribution of the events in 
the energy of neutrinos at the interaction point)
is shown in Figure \ref{fig1} for five cases:  
1.~a fully `idealized' case;
$2.-4.$~the three cases when oscillations, absorption and 
live-time are considered once at the time; 
5.~the case when all three effects are included. 
The typical  energies are in the range 1-200 TeV,
after inclusion of absorption (that produces a downward shift).
For $E_\nu=50$~GeV$-$1~PeV, these effects reduce
the number of events by:
\begin{equation}
\mbox{abs.: }0.81,
\mbox{  liv.: }0.78,
\mbox{  osc.: }0.51 
\end{equation}
The impact of all these effects, and in particular the one of oscillations,
are rather important. In particular, the number of events 
expected for case 1.\ is:
\begin{equation}
N_{\mu+\bar{\mu}}^{ideal}=29.1
\label{idresult}
\end{equation}
This number is compatible with the 
41 events found in \cite{amh} (see fig.\ 1 there), 
when we consider that in \cite{amh}
the spectral index is assumed to be very hard, 
$\Gamma=2$. But after the inclusion of 
oscillations, absorption and live-time,
the decrease is much stronger:
\begin{equation}
N_{\mu+\bar{\mu}}=9.3
\label{result}
\end{equation}
This is the main result of the present work.

\begin{figure}[t]
\includegraphics[width=.44\textwidth,angle=270]{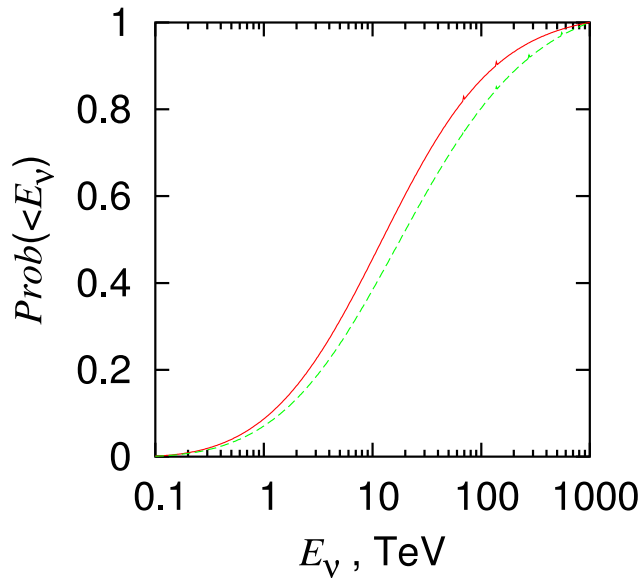}
\caption{\em Cumulative distribution of the number of 
muons + antimuons above $E_{th}=50$~GeV.
The curve displaced at high energies is the `idealized' case
(corresponds to the higher curve of fig.\ref{fig1})
the other one includes the effects of 
absorption, live-time and 3 flavor oscillations
(corresponds to the lower curve of fig.\ref{fig1}).
\label{fig2}}
\end{figure}

We present cumulative 
curves in Figure \ref{fig2}. These permit to rescale the above results 
if the maximal energy of the neutrino spectrum (=the cut of
the power spectrum) is lower than the value 
we allowed, $E_{\nu, max}=1$~PeV. For instance, 
if we limit ourself to what we know from photons,
we could believe that the cut in the neutrino energy 
happens as early as at 5 TeV. This would mean a dramatic
reduction factor of 0.31 to be applied to 
the number in eq.~(\ref{result}).
Instead, in the more realistic case when the 
proton spectrum is cut at the energy of the 
knee ($E_{p,max}=3$~PeV), one expects a cut
for the neutrino spectrum somewhere close to 250 TeV  (and 
close to 0.5~PeV for gamma spectrum). 
This means that the number in  eq.~(\ref{result}) should 
be diminished, but only by~5~\%.

Finally, let us note that the numbers  of
eqs.~(\ref{idresult}) and (\ref{result})  apply to 
a detector with unit efficiency of detection.\footnote{In real detectors,
the efficiency is usually included in the ``effective area'', 
that is an increasing function of the energy.}


\section{Summary and discussion\label{dis}}
The recent H.E.S.S.~measurements support the view that 
RX J1713-3946 is a source of neutrinos with energies 
at TeV and above.
Existing data already permit to predict the neutrino flux
to a reasonable level of approximation.
Future gamma-ray data should clarify the picture,  
and possibly reveal the extension of the power spectrum.

We calculated the expected number 
and distribution of neutrino events
in underwater neutrino telescopes from RX J1713.7-3946.
These calculations cannot be considered
definitive for a number of reasons
(e.g., CR are assumed to be solely protons,
a power law spectrum is assumed, 
`neutrino regeneration' is treated
in the simplest approximation, only a perfect 
detector is considered). 
Also, we did not attempt to estimate the background, though 
this was done purposely:
we believe that it should be estimated 
during detector operation,
and we are aware of a number of theoretical uncertainties 
(generally on high energy
part of atmospheric neutrinos flux~\cite{giu},
and more specifically on the prompt contribution).

However,  
we improved over the existing calculation 
of the neutrino signal from RX J1713-3946 \cite{amh} 
in several senses:
we considered a deviation from
strict equality $\Gamma=2$, we treated the 
neutrino interactions at NLO, we estimated absorption in the Earth
and live-time of data acquisition, and most importantly,
we included 3 flavor oscillations. 
Our calculations, in particular eq.~(\ref{result}),
suggest that a detector located in the Northern hemisphere
should have an effective area of $\sim {\rm km^2}$ 
and/or a long data taking time in order to see RX J1713-3946 
as a source of high-energy neutrinos.

\vskip1mm
\noindent We thank for pleasant discussions and help 
V.~Berezinsky, 
A.~Butkevich,
M.~Cirelli,
P.~Desiati,
S.~Dugad,
W.~Fulgione,
P.~Ghia,
T.~Montaruli,
G.~Navarra,
I.~Sokalsky,
A.~Strumia,
R.~Thorne,
Y.~Uchiyama.

\end{twocolumn}

\begin{thebibliography}{99}
\bibitem{mrstnlo}
The needed {\tt FORTRAN} programs are in 
{\tt mrst02cc.f} 
(trivially modified), and can be found at 
the web site durpdg.dur.ac.uk/hepdata/mrs.html
\end{thebibliography}
\end{document}